\documentstyle[12pt, epsf, psfig]{article}
\def\reference{\parskip 0pt\par\noindent\hangindent 0.5 truecm}

\begin{document}
%
%
\title{The HI Parkes Zone of Avoidance Shallow Survey}
%


\author{P. A. Henning $^{1}$ \and
 L. Staveley-Smith $^{2}$ \and
 R. C. Kraan-Korteweg $^{3}$ \and
 E. M. Sadler $^{4}$
} 

\date{}
\maketitle

{\center
$^1$ Institute for Astrophysics, University of New Mexico,
Albuquerque, NM, 87131, USA\\henning@tinsley.phys.unm.edu\\[3mm]
$^2$ Australia Telescope National Facility, CSIRO, P. O. Box 76, 
Epping, NSW 1710,
Australia\\Lister.Staveley-Smith@atnf.csiro.au\\[3mm]
$^3$ Departamento de Astronomia, Universidad de Guanajuato, Guanajuato,
Gto. CP 36000, Mexico\\kraan@astro.ugto.mx\\[3mm]
$^4$ School of Physics, University of Sydney, NSW 2006, Australia\\
EMS@astrop.physics.usyd.edu.au\\[3mm]
}

\begin{abstract}
The HI Parkes Zone of Avoidance Survey is a 21 cm blind search
with the multibeam receiver on the 64-m radiotelescope, looking
for galaxies hidden behind the southern Milky Way.
The first, shallow (15 mJy rms) phase of the survey has uncovered 107 galaxies,
two-thirds of which were previously unknown.
The addition of these galaxies to existing extragalactic
catalogs allows the connectivity of a very long, thin filament
across the Zone of Avoidance within 
$3500 \,\rm\,km\,s^{-1}$ to become evident.
No local, hidden, very massive objects were uncovered.
With similar results in the north (The Dwingeloo Obscured Galaxies
Survey)
our census of the most dynamically important HI-rich nearby
galaxies is now complete, at least for those objects whose HI
profiles are not totally buried in the Galactic HI signal.
Tests are being devised to better quantify this remaining ZOA for blind HI
searches.
The full survey is ongoing, and is expected to produce
a catalog of thousands of objects when it is finished.

\end{abstract}


\bigskip

%
%

\section{Introduction}

The dust and high stellar density of the Milky Way obscures
up to 25\% of the optical extragalactic sky,
creating a Zone of Avoidance (ZOA).
The resulting incomplete coverage of surveys of external
galaxies leaves open the possibility that dynamically important
structures, or even nearby massive galaxies, remain undiscovered.
 
Careful searches in the optical and infrared wave bands can narrow the ZOA,
(see Kraan-Korteweg, this volume)
but in the regions of highest obscuration and infrared confusion, only
radio surveys can find galaxies.
The 21 cm line of neutral hydrogen (HI) passes readily through the obscuration,
so galaxies with sufficient HI can be found through detection of their 21 cm
emission.  
Of course, this method will miss HI-poor, early-type galaxies, and cannot 
discriminate
HI galaxies with redshifts near zero velocity from Galactic HI.

Here we describe an HI blind survey for galaxies in the southern ZOA 
conducted with the new multibeam receiver on the 64-m Parkes telescope.
A survey of HI galaxies in the northern ZOA is underway with the Dwingeloo
radiotelescope (Henning et al. 1998; Rivers et al. this volume).

\section{The Shallow Survey}

\subsection{Observing Strategy}

The HI Parkes ZOA survey covers the southern ZOA 
($212\deg \le l \le 36\deg; \vert b \vert \le 5\deg$) over the
velocity range (cz) = $-1200$ to $12700\,\rm\,km\,s^{-1}$.
The multibeam receiver is a focal plane array with 13 beams arranged in an
hexagonal grid.
The spacing between adjacent beams is about two beamwidths, each beamwidth
being 14 arcmin.
The survey is comprised of 23 contiguous rectangular fields 
which are scanned parallel to the galactic equator.
Eventually, each patch will be observed 25 times, with scans offset by
about 1.5 arcmin in latitude.
The shallow survey discussed here consists of two scans in
longitude separated by 
$\Delta$b = 17 arcmin, resulting in an rms noise of about 15 mJy, equivalent 
to a 5$\sigma$ HI mass detection limit of $4 \times 10^6$ d$^2\!\!_{\rm Mpc}$
M$_{\odot}$ (for a galaxy with the typical linewidth of 
$200{\,\rm\,km\,s^{-1}}$).

\subsection{Data Visualization}

After calibration, baseline-subtraction, and creation of data cubes,
all done with specially developed routines based on aips++
(Barnes et al. 1998, Barnes 1998)
the data are examined by eye using the visualization
package Karma
(http://www.atnf.csiro.au/karma/).
The data are first displayed as right ascension -- velocity
planes, in strips of constant declination.
Data cubes are
then rotated, and right ascension -- declination planes are
checked for any suspected galaxies (eg. Figure 1).

 \begin{figure}
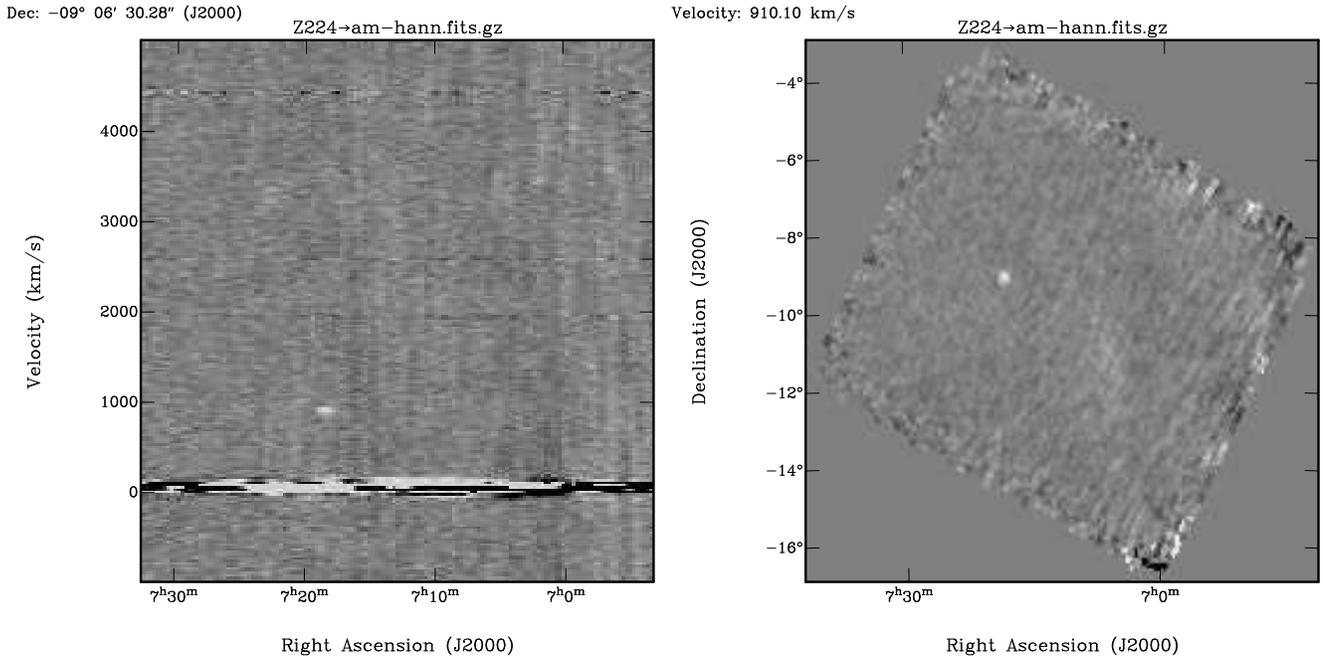

 \begin{center}
 \hbox{
 \psfig{file=fig1a,height=8.7cm}
 \psfig{file=fig1b,height=8.7cm}
 }
\caption{Left panel shows a right ascension -- velocity slice to
5000${\,\rm\,km\,s^{-1}}$ which includes a galaxy
discovered by the survey.
Galactic HI appears as the strong horizontal feature at zero velocity.
Note the extragalactic HI signal at $7^{\rm h}18^{\rm m}$, 
$900{\,\rm\,km\,s^{-1}}$.
Right panel shows a right ascension -- declination plane at the
velocity of the suspected signal.
The galaxy is evident at $7^{\rm h}18^{\rm m}$, $-9\deg$.}
  \label{fig-1}            
  \end{center}
  \end{figure}

\section{Galaxies Found by the Shallow Survey}

The shallow 21-cm survey of the southern ZOA 
has been completed, and 107 galaxies with peak HI flux densities $\geq$ 
about 80 mJy 
have been cataloged.
Refinement of the measurement of their HI characteristics is ongoing,
but the objects
seem to be normal galaxies.
However, of three large multibeam ZOA galaxies imaged in HI with the ATCA,
two were seen to break up into complexes of HI suggestive of
tidally-interacting systems (Staveley-Smith et al. 1998)
Continued follow-up synthesis observations are planned to investigate
the frequency of these interacting systems in this purely HI-selected
sample.

Most of the galaxies are within $4000{\,\rm\,km\,s^{-1}}$,
which is about the redshift limit 
for detection of normal spirals
of this shallow phase of the survey.
As the deep survey continues, spirals at higher velocities will be
recovered.
The effective depth of the shallow survey
is not quite sufficient to recover large numbers
of galaxies which might be associated with the Great Attractor
(but see Juraszek et al. this volume.)
However, a striking feature becomes apparent with the addition of
the ZOA galaxies.
An enormous filament, which crosses the ZOA twice, is clearly evident
when these ZOA data are displayed along with optically-known galaxies
above and below the plane within $3500{\,\rm\,km\,s^{-1}}$ (Fig. 2.)
This structure snakes over $\sim180\deg$ through the southern sky.
Taking a mean distance of $30{h^{-1}}$ Mpc, this implies a linear
size of $\sim100{h^{-1}}$ Mpc, with thickness of $\sim5{h^{-1}}$ Mpc
or less.

Also, note the relative emptiness of the Local Void.
Three hidden galaxies found
on a boundary of the Void (l $\sim30\deg$) lie at
$\sim 1500 {\,\rm\,km\,s^{-1}}$.  Two of these objects were also recovered
by the Dwingeloo Obscured Galaxies Survey (Rivers et al. this volume.)
The positions and redshifts of these objects are consistent with their
being members of the cluster at this location proposed by Roman et al. (1998).

  \begin{figure}
  \begin{center}
  \psfig{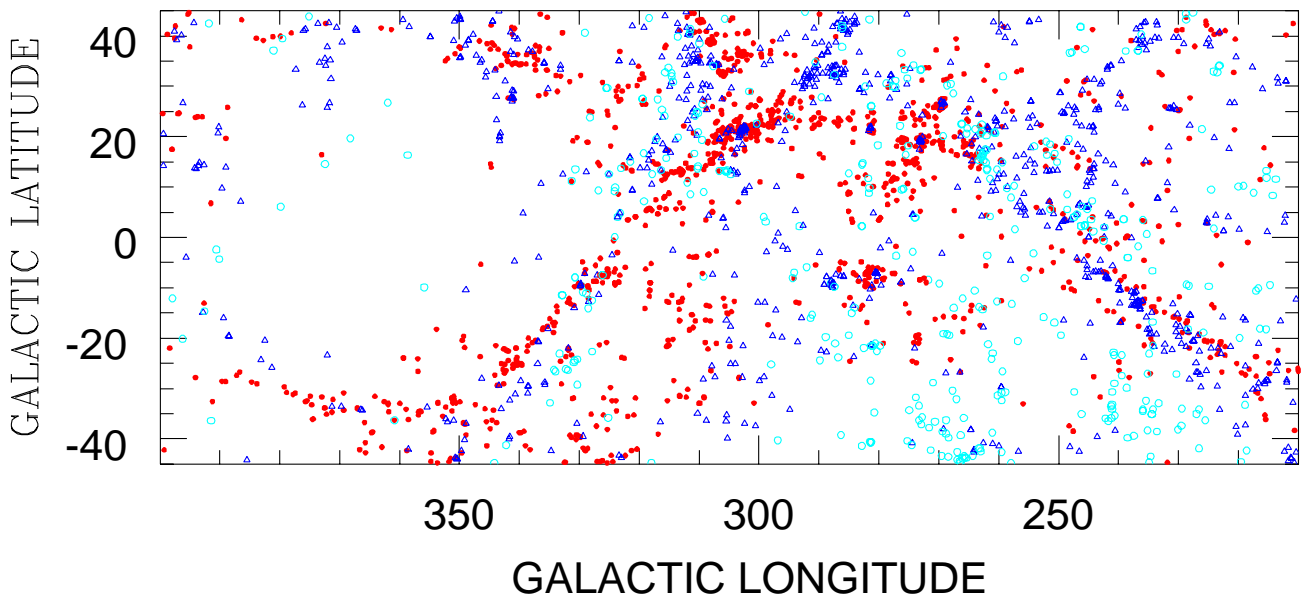}
\caption{Galaxies within v $\leq3500{\,\rm\,km\,s^{-1}}$.
The open circles mark galaxies between $500 - 1500{\,\rm\,km\,s^{-1}}$,
triangles for $1500 - 2500{\,\rm\,km\,s^{-1}}$, and dots for
those with $2500 - 3500{\,\rm\,km\,s^{-1}}$.
High latitude data are taken from the literature (LEDA).
Also plotted are galaxies from deep optical ZOA surveys
(Kraan-Korteweg et al. this volume.)  The 
galaxies discovered by the multibeam shallow ZOA survey
fill in the lowest galactic latitudes, where optical surveys
fail.  
}
  \label{fig-2}
  \end{center}
  \end{figure}

Of the 107 objects found,
28 have counterparts in the NASA/IPAC Extragalactic Database (NED)
with matching positions and redshifts.
Optical absorption,
estimated from the Galactic dust data of
Schlegel et al. (1998), ranges from A$_{\rm B}$ = 1 to more than 60 mag
at the positions of the 107 galaxies,
and is patchy over the survey area.
No objects lying behind more than about 6 mag of obscuration have
confirmed counterparts in NED, as expected.

The shallow multibeam HI survey connects structures all the way across
the ZOA within $3500{\,\rm\,km\,s^{-1}}$ for the first time.
The ongoing, deep ZOA survey will have sufficient sensitivity to
connect structures at higher redshifts.

While 14 of the 107 galaxies lie within 
$1000{\,\rm\,km\,s^{-1}}$ and are therefore fairly nearby,
all of the newly-discovered objects have peak HI flux densities an order
of magnitude or more lower than the Circinus galaxy.
Thus, it seems our census of the most dynamically important, HI-rich nearby
galaxies is now complete, at least for those objects with velocities
offset from Galactic HI.
Simulations are currently being devised to investigate our sensitivity
to HI galaxies whose signals lie within the frequency range of the Milky
Way's HI.
This will be done by embedding artificial HI signals of varying strength,
width, position, and frequency, into real data cubes.
Then, an experienced HI galaxy finder (PH) will examine the cubes without
previous knowledge of the locations of the fake galaxies.
In this way, we hope to quantify better this remaining blind spot of the
HI search method.

\section*{Acknowledgements}

We thank HIPASS ZOA collaborators
R. D. Ekers, A. J. Green, R. F. Haynes, S. Juraszek, M. J. Kesteven,
B. S. Koribalski, R. M. Price, and
A. Schr\"oder.
This research has made use of the NASA/IPAC Extragalactic Database (NED)
which is operated by the Jet Propulsion Laboratory, Caltech, under
contract with the National Aeronautics and Space Administration.
We have also made use of the Lyon-Meudon Extragalactic Database (LEDA),
supplied by the LEDA team at the Centre de Recherche Astronomique de
Lyon, Observatoire de Lyon.
The research of P.H. is supported by NSF Faculty Early Career Development
(CAREER) Program award AST 95-02268.

\section*{References}

\reference Barnes, D.G. 1998, in ADASS VII, eds. Albrecht, R., Hook, R.N.,
    \& Bushouse, H.A., San Francisco:  ASP
\reference Barnes, D.G., Staveley-Smith, L, Ye, T., \& Osterloo, T.
    1998, in ADASS VII, eds. Albrecht, R., Hook, R.N.,
    \& Bushouse, H.A., San Francisco:  ASP
\reference Henning, P.A., Kraan-Korteweg, R.C., Rivers, A.J., Loan, A.J.,
    Lahav, O., \& Burton, W.B. 1998, AJ, 115, 584
\reference Roman, A.T., Takeuchi, T.T., Nakanishi, K., \& Saito, M.
    1998, PASJ, 50, 47
\reference Schlegel, D.J., Finkbeiner, D.P., \& Davis, M. 1998, ApJ, 500,
    525
\reference Staveley-Smith, L., Juraszek, S., Koribalski, B.S., Ekers,
    R.D., Green, A.J., Haynes, R.F., Henning, P.A., Kesteven, M.J.,
    Kraan-Korteweg, R.C., Price, R.M., \& Sadler, E.M. 1998, AJ, in press.

\end{document}